\documentclass{sigchi-ext}
\usepackage[T1]{fontenc}
\usepackage{textcomp}
 \usepackage{amsmath}
\usepackage[scaled=.92]{helvet} % for proper fonts
\usepackage{graphicx} % for EPS use the graphics package instead
\usepackage{balance}  % for useful for balancing the last columns
\usepackage{booktabs} % for pretty table rules
\usepackage{ccicons}  % for Creative Commons citation icons
\usepackage{ragged2e} % for tighter hyphenation
\usepackage{subfigure}
\usepackage[noadjust]{cite}
\usepackage{url}
\usepackage{multirow}
\usepackage{color}
\usepackage{refcount}
 \usepackage{enumitem}
 
\def\plaintitle{SIGCHI Extended Abstracts Sample File: Note Initial
  Caps} 
\def\emptyauthor{}
\def\plainkeywords{Authors' choice; of terms; separated; by
  semicolons; include commas, within terms only; required.}

\title{Understanding Book Popularity \\ on Goodreads}

\numberofauthors{5}
\author{
\alignauthor{\textbf{Suman Kalyan Maity}\\
       \affaddr{Dept. of CSE}\\
       \affaddr{IIT Kharagpur, India}
       \email{sumankalyan.maity@\\cse.iitkgp.ernet.in}}
       \alignauthor{\textbf{Vishnu Choudhary}\\
       \affaddr{Dept. of CSE}\\
       \affaddr{IIT Kharagpur, India}
       \email{bansal.jhs@gmail.com}}
       \vfil \alignauthor{\textbf{Ayush Kumar}\\
       \affaddr{Dept. of CSE}\\
       \affaddr{IIT Kharagpur, India}
       \email{ayush235317@gmail.com}}
       \alignauthor{\textbf{Animesh Mukherjee}\\
       \affaddr{Dept. of CSE}\\
       \affaddr{IIT Kharagpur, India}
       \email{animeshm@cse.iitkgp.ernet.in}}
       \vfil \alignauthor{\textbf{Ankan Mullick}\\
       \affaddr{Bing}\\
       \affaddr{Microsoft India}
       \email{ankan.mullick@microsoft.com}}}

\definecolor{linkColor}{RGB}{6,125,233}
\hypersetup{%
  pdftitle={\plaintitle},
  pdfauthor={\emptyauthor},
  pdfkeywords={\plainkeywords},
  bookmarksnumbered,
  pdfstartview={FitH},
  colorlinks,
  citecolor=black,
  filecolor=black,
  linkcolor=black,
  urlcolor=linkColor,
  breaklinks=true,
}

\begin{document}

%% For the camera ready, use the commands provided by the ACM in the Permission Release Form.
\CopyrightYear{2018}
\setcopyright{rightsretained}
\conferenceinfo{GROUP '18}{January 7--10, 2018, Sanibel Island, FL,
USA}\isbn{978-1-4503-5562-9/18/01}
\doi{https://doi.org/10.1145/3148330.3154512}
\copyrightinfo{\acmcopyright}

\maketitle

% Uncomment to disable hyphenation (not recommended)
% https://twitter.com/anjirokhan/status/546046683331973120
\RaggedRight{} 

% Do not change the page size or page settings.

\begin{abstract}
Goodreads has launched the Readers Choice Awards since 2009 where users are able to nominate/vote books of their choice, released in the given year. In this work, we question if the number of votes that a book would receive (aka the popularity of the book) can be predicted based on the characteristics of various entities on Goodreads. \iffalse We propose a prediction framework to predict book popularity (no. of votes a book receives) taking cue from various characteristic properties of the Goodreads entities.\fi We are successful in predicting the popularity of the books with high prediction accuracy (correlation coefficient $\sim$0.61) and low RMSE ($\sim$1.25). User engagement and author's prestige are found to be crucial factors for book popularity.
\end{abstract}
\keywords{book popularity; Goodreads; prediction}
% 
% \category{H.5.m}{Information interfaces and presentation (e.g.,
%   HCI)}{Miscellaneous}\category{See}{\url{http://acm.org/about/class/1998/}}{for
%   full list of ACM classifiers. This section is required.}
\category{H.4.m}{Information Systems Applications}{Miscellaneous}
%A category including the fourth, optional field follows...
\category{J.4}{Computer Applications}[Social and Behavioral Sciences]
\category{K.4.2}{Computers And Society}[Social Issues]
\section{Introduction}
Popularity/success of a book is important not only for the authors but also for the publishers, the professional book reviewers and the book selling platforms like Amazon, eBay etc. The initial popularity of books can have a significant impact how the eventual sales would be~\cite{suman}. Understanding book popularity is a difficult task even for an expert working in the publication industries. Some of the eventual award winners and best sellers have gone through one or multiple rejections before being finally accepted by a certain publisher. There are potentially many influencing factors that can impact popularity of a book. Broadly these factors can be categorized into (i) intrinsic and (ii) extrinsic factors. Intrinsic factors correspond to the content and the quality of the book like how interesting the book is, how engaging the story line is, novelty, style of writing etc. However, these quality factors of the books are very different for different genres.
\marginpar{%
 \fbox{%
    \begin{minipage}{0.92\marginparwidth}
\textbf{Related Work}\\
Predicting popularity of content on social media has been widely researched. Most of these studies focus on popularity prediction of various Twitter specific entities. Extensively studied among these being the hashtags. Tsur and Rappoport~\cite{what} studied popularity of hashtags based on content features of tweets only. Ma et al. in~\cite{maon} proposed a framework for predicting popularity of newly emerging hashtags. Kong et al.~\cite{kong} studied the burstiness (sudden rise in hashtag usage and quick fall thereafter) of hashtag on a temporal scale. Maity et al.~\cite{merge} studied various factors affecting the popularity of hashtag compounds (two or more hashtags merging together). In a recent work by Maity et al.~\cite{suman}, the authors have studied how book reading behavior on Goodreads can determine Amazon Best Sellers.\\

\end{minipage}}\label{rel_work}}
For example, a successful thriller requires a credible story-line, complex twists and plots, escalating stakes and tension whereas a popular romantic novel demands demonstration of strong and healthy relationship, sexual tension, happy and optimistic endings  etc. A great mystery novel involves secrets, misdirection of clues, a relatable protagonist etc.~\cite{hall}. Therefore, finding common grounds is difficult for quantification of the quality aspects for different genres. On the other hand, extrinsic factors include the readers' reading behaviors, social contexts, reviews by the critics etc. which are relatively easier to obtain. In this paper, we shall consider the readers' reading, reviewing characteristics for the books, engagement activities, authors' prestige etc. to understand book popularity in Goodreads. Specifically, we are interested in predicting book votes as a popularity metric.

Goodreads is a community-driven social cataloging site which has grown exponentially into one of the most popular social book reading and recommendation sites. Goodreads provides various opportunities like quizzes, trivia apart from social book reading so as to engage their users. Goodreads Readers Choice Award is such an attempt. This was first launched in 2009. From then on, Goodreads users can take part in deciding the recipients of this award by nominating as well as by voting for their nominations. There are 20 categories like Fiction, Thrillers, Fantasy, Romance etc. of awards and in each category 15 official nominations are made. In the first round of voting, users can nominate books to be included in the awards as write-in candidates; five in each category get added to the group of official nominees, making the total 20 in each category. Using APIs and automated crawls, we gathered all the books' data prior to the start of the voting phase. In table~\ref{tabaward}, we show all the award categories.  \iffalse In this paper, we focus on understanding the popularity of books (votes that they receive).\fi 

\textbf{Present work:} In this study, we aim to determine the salient factors from various Goodreads entities that contribute to the popularity of books in terms of the number of votes they receive. Towards this objective, we have considered collective user engagement behavior and show that this is an important aspect to understand book popularity alongside author's prestige. The user engagements toward books characterized by the rating/review behavior, shelve characterization and organization - a very unique utility that the Goodreads platform provides to its users, are important determinant. Further, we observe that author's prestige features like -- the avg. rating of an author, the number of awards received by an author of a book etc. are crucial to the book popularity.

% \begin{margintable}[1pc]
% 
%   \begin{minipage}{\marginparwidth}
%     \centering
%     \vspace{-50mm}
% \small
% \captionsetup{font=small}
% \caption{Award Categories}
% \label{tabaward}
% \resizebox{4cm}{!}{\begin{tabular}{|c|c|}\hline
% Fiction 
% History \& Biography \\\hline
% Mystery \& Thriller\\\hline
% Science \& Technology\\\hline
% Historical Fiction\\\hline
% Food \& Cookbooks\\\hline
% Fantasy \\\hline
% Graphic Novels \& Comics \\\hline
% Romance\\\hline
% Poetry\\\hline
% Science Fiction \\\hline
% Debut Goodreads Author\\\hline
% Horror\\\hline
% Young Adult Fiction\\\hline
% Humor \\\hline
% Young Adult Fantasy\\\hline
% NonFiction \\\hline
% Middle Grade \& Children's\\\hline
% Memoir \& Autobiography \\\hline
% Picture Books\\\hline
% \end{tabular}}
% \end{minipage}
% \end{margintable}

\begin{table}[h]
\begin{center}
\small
 \captionsetup{font=small}
\caption{Award Categories}
\label{tabaward}
\resizebox{6cm}{!}{\begin{tabular}{|c|c|}\hline
Fiction & History \& Biography \\\hline
Mystery \& Thriller& Science \& Technology\\\hline
Historical Fiction& Food \& Cookbooks\\\hline
Fantasy & Graphic Novels \& Comics \\\hline
Romance& Poetry\\\hline
Science Fiction & Debut Goodreads Author\\\hline
Horror & Young Adult Fiction\\\hline
Humor & Young Adult Fantasy\\\hline
NonFiction & Middle Grade \& Children's\\\hline
Memoir \& Autobiography & Picture Books\\\hline
\end{tabular}}
\end{center}
\end{table}

% Apart from Twitter, there have also been some studies on popularity of entities on other social media. Maity et al.~\cite{suman} studied topic popularity on Quora while Ge et al.~\cite{fiverr} studied gig quality on Fiverr. To the best of our knowledge, this is the first study which looks into the popularity of books which is quite different from other social media entities. 
\section{Factors behind Book Popularity:} In this section, we try to understand various factors driving book popularity. We consider two major types of factors:\\
\marginpar{%
 \fbox{%
    \begin{minipage}{0.93\marginparwidth}
\textbf{Discriminative features:}
\iffalse Here, we discuss the discriminative power of the individual features used in the prediction model. \fi We use $RELIEFF$ feature selection algorithm~\cite{relief} to rank the attributes. In table~\ref{feature_rank}, we show the rank of the features in terms of their discriminative power for prediction. The rank order clearly indicates the dominance of the user engagement features, however author's rating seems to be the most discriminative feature. In the top 10, three of the author features find place. Among engagement features, shelve diversity, no. of books in `read' or `currently-reading' shelve are the important factors for popularity prediction. This suggest that Goodreads' user engagement is crucial factor for popularity prediction. Among author features, apart from rating of the author, other prestige features of the authors like no. of awards received, no. of rating received, no. of best sellers act as prominent popularity prediction features.
\end{minipage}}\label{discrim} }
User engagement towards books - these includes various user engagement characteristics toward books like rating, reviewing, organizing in shelves etc.\\
Author characteristics - these factors are related to the prestige of an author of a nominated book. 
Subsequently we shall use them as features to our prediction model.
\iffalse Formally, we attempt to learn a target function $f(X) = \log(v)$, where $X$ is the feature vector of a given book $b$ and $v$ is the no. of votes received by the nominated book $b$ after the voting gets closed. We are interested in
predicting the magnitude of the popularity, thus while a book with 100 votes is very different from a book with 1000 votes, 30000 is similar to 31000. Taking logarithm captures this observation. We further categorize the books by the award category. For each of these categories, we separately learn the prediction model from the books leaving all the books from that particular category out of the learning module and find the prediction accuracies by considering the books of that category as the test set.\fi
\subsection*{\textbf{User Engagement:}} These factors are extracted from various characteristic properties of Goodreads' user engagement toward books which we have already studied earlier.
\begin{itemize}[noitemsep,nolistsep]
 \item Average rating of the book given by the users.
\item Number of ratings given by the users.
\item Number of 4-star ratings given to the book.
\item Number of 5-star ratings given to the book.
\item User rating entropy of the book.
\item Number of reviews received by the users.
\item Review Sentiments - We take only 30 reviews of the books and use NLTK sentiment analysis tool\footnote{\url{http://text-processing.com/demo/sentiment/}} to find sentiment of the reviews. 
We then take the avg. positive sentiment score, negative sentiment score, neutral sentiment score and standard deviation of these sentiment scores as features to our model. In total, we have six set of features here.
\item Number of genres of the book. 
\item Number of different user shelves the book is present in.
\item Shelve diversity of the book - Similar to rating entropy, we calculate shelve diversity. Formally, shelve diversity (ShelveDiv) is defined as follows:
\begin{equation*}
 ShelveDiv(b) = -\sum_{j \in shelve_set}s_j\times\log(s_j)
 \vspace{-3mm}
\end{equation*}
where $s_j$ is the probability that the book belongs to the $j^{th}$ user shelve in the set of user bookshelves.
\item Number of users who have added the book in `currently reading' or `read' shelves.
\item Number of users who have added the book in `to read' shelves.
\end{itemize}
\vspace{-3mm}
\subsection*{\textbf{Author Characteristics:}} These factors are extracted from various characteristic properties (mostly prestige) of the authors of the nominated book.

\vspace{-3mm}
\begin{itemize}[noitemsep,nolistsep]
\item Number of books written by the author of the book.
\item Average rating of the author of the book.
\item Number of ratings received by the author of the book.
\item Number of reviews received by the author of the book.
\item Number of distinct awards received by the author.
\item Number of `best seller' books of the author.
\item Follower count of the author.
\item Number of common shelves among the authors' books.
\item Number of unique shelves among the authors' shelves.
\end{itemize}

\begin{table}[h]
\vspace{-3mm}
\small
\centering
\caption{\vspace{-3mm}Top 10 predictive features.}
  \label{feature_rank}
  \resizebox{8cm}{!}{
 \begin{tabular}{ |l|l|}
\hline
Rank & Features  \\ \hline
1&Avg. rating of the author  \\
 2& Shelve Diversity  \\
 3 & No. of users who have added the book in `currently reading' or `read' shelves \\
 4 & No. of awards received by the author\\
 5 & No. of different shelves the book is present in \\
 6 & No. of users who have added the book in `to-read' shelves\\
 7 & Rating Entropy of the book\\
 8 & No. of reviews received by the book \\
 9 & No. of 5-star ratings received by the book \\
 10 & No. of ratings received by the author \\\hline
%  11 & Avg. rating score of the book & Book \\
%  12 & No. of ratings received by the book & Book \\
%  13 & No. of best sellers of the author & Author\\
%  14 & No. of 5-star ratings received by the book & Book\\
%  15 & No. of books written by the author & Author\\\hline
  \end{tabular}}
%   \vspace{-1mm}
\end{table}
% \vspace{-3mm}
\section{Prediction Framework} We shall now use the above user engagement and author characteristics as features for our prediction model. 
\begin{margintable}[1pc]
  \begin{minipage}{\marginparwidth}
    \centering   
%     \vspace{-5mm}
\small
\captionsetup{font=small}
\caption{Prediction results for the highly voted books}
  \label{tabhigh}
 \resizebox{4.5cm}{!}{\begin{tabular}{ |c|c|c| }
\hline
 & Corr. Coeff.($r$) & RMSE  \\ \hline
top 100 & 0.66 & 2.72 \\\hline
top 50 & 0.63& 1.56 \\\hline
top 20&  0.75& 1.15 \\\hline
top 10 & 0.68 & 1.04 \\\hline
top 5& 0.7& 1.17\\\hline
top 3 & 0.83& 2.2\\\hline
  \end{tabular}}
\end{minipage}
\end{margintable}
\begin{margintable}[1pc]

  \begin{minipage}{\marginparwidth}
    \centering
%     \vspace{-2mm}
\small
\captionsetup{font=small}
\caption{Prediction results for the Award category specific prediction}
  \label{tabcat}
 \resizebox{4.5cm}{!}{\begin{tabular}{ |c|c|c| }
\hline
Category & $r$ & RMSE  \\ \hline
Fiction & 0.76 & 0.83 \\\hline
Mystery \& Thriller& 0.69 & 1.08\\\hline
Historical Fiction&  0.84& 1.15 \\\hline
Fantasy & 0.82 & 0.89 \\\hline
Romance& 0.49& 1.63\\\hline
Science Fiction & 0.67& 1.19\\\hline
Horror & 0.73& 1.11\\\hline
Humor & 0.7& 1.25\\\hline
NonFiction & 0.7& 1.33\\\hline
Memoir \& Autobiography &0.48 & 1.18\\\hline
History \& Biography & 0.66& 1.08\\\hline
Science \& Technology & 0.41& 1.3\\\hline
Food \& Cookbooks & 0.61& 1.77\\\hline
Graphic Novels \& Comics & 0.39& 1.39 \\\hline
Poetry& 0.52&1.59 \\\hline
Debut Goodreads Author & 0.87& 0.82\\\hline
 Young Adult Fiction & 0.8& 0.86\\\hline
 Young Adult Fantasy &0.79 & 1.46\\\hline
  Middle Grade \& Children's &0.61 & 1.2\\\hline
  Picture Books &0.66 & 1.38\\\hline
\end{tabular}}
\end{minipage}
\end{margintable}
We consider 400 books from all the 20 award categories of Goodreads Choice Award in 2015 for our prediction task. We perform a 10-fold cross-validation on the data sample. We use Support Vector Regression (SVR) for the prediction. For evaluating how good the prediction is, we use Pearson correlation coefficient ($r$) and root mean square error (RMSE). We achieve high correlation coefficient ($\sim0.61$) and low root mean square error ($\sim1.25$). We observe that user engagement feature type is the strongest feature type contributing to correlation coefficient of $0.59$ with RMSE value of $1.29$ whereas with only author features, we achieve corr. coeff. of $0.44$ and RMSE value of $1.41$.

\subsection*{\textbf{Predicting the votes of the highly voted books:}} 
Apart from overall prediction of the votes for the book nominations, we also investigate how our model performs in predicting the votes of the highly voted books. In specific, we ask if the features are able to suitably discriminate these books and if the predictions for them are better or worse than the overall prediction. \iffalse In order to answer these questions, we separately perform predictions for top 3, 5, 10, $\dots$ 100 most voted books.In table~\ref{tabhigh}, we present the prediction results. Note that while predicting the top 3, 5, 10 books' vote, we separately train our model on the remaining 397, 395, 390 books respectively. \fi We observe that our prediction model can very well predict the highly voted books' vote and the correlation coefficient is always higher than the overall case (see table~\ref{tabhigh}). For predicting the votes of the top 3 most voted  books, our model achieves a very high correlation coefficient of $0.83$ (although the RMSE value goes a bit higher).

% \vspace{-2mm}
% \vspace{-2mm}
\subsection*{\textbf{Award category specific prediction:}} We further categorize the books into the Award categories to investigate whether such categorization helps in improving the prediction accuracy. For prediction of votes of the books in each category, we train the prediction model on all the books except for the books belonging to that category. The set of books in this category acts as a test set for the prediction task. In table~\ref{tabcat}, we show the results of the prediction. In most of the categories, we observe significant improvement in prediction accuracy from the case with no categorization. We also observe that in some award categories, prediction accuracies fall e.g., Romance, Science \& Technology, Memoir \& Autobiography, Graphic Novels \& Comics etc. 

\section{Conclusions and Implications}
In summary, we propose a framework for predicting popularity (no. of votes) of books. Our proposed model achieves a high ccorrelation coefficient $\sim$0.61 with low RMSE ($\sim$1.25). We observe that the user engagement features are the most discriminative ones compared to the others. Our prediction framework can predict votes of the highly voted books with higher accuracy than the above base case. The stratification of the books into award categories further enhances the prediction accuracies for most of the categories significantly.

Our research has important implications. It shows that initial rating, reviewing, user engagement features obtained from collective Goodreads user behavior along with authors' prestige can efficiently determine popularity of books. Our proposed system can early predict book popularity using these above features which can be easily obtainable. This early prediction can be effective in several ways - (i) act as guide for recommending appropriate books to the new users joining Goodreads and, (ii) help the book selling platforms like Amazon, eBay by forecasting early the eventual fate of a book/group of books/genre so that these platforms are able to launch proper and focused advertisements/promotional campaigns to boost up the sales.
\bibliographystyle{SIGCHI-Reference-Format}
\bibliography{ref}

\end{document}